\documentclass[%
 reprint,
 amsmath,amssymb,
 aps,
]{revtex4-2}

\usepackage{graphicx}
\usepackage{dcolumn}
\usepackage{bm}
\usepackage{xcolor} 
\usepackage{csquotes}

\usepackage{physics}

\begin{document}

\preprint{APS/123-QED}

\title{Probing the Quantum and Classical Boundary: A Tabletop Experiment Using Quantum Optics}
\author{Muchen He}
\affiliation{Guangzhou, China}
\author{Jizhe Lai}
\affiliation{Guangzhou, China}

\date{\today}

\begin{abstract}
In this work, we propose a simple but effective experiment for probing the boundary in which a wave-function collapses. Using a quantum optics system interacting with a photomultiplier tube (PMT), one is able to determine the number of electrons needed to interfere with the \enquote{which-path} information to cause the collapse of a quantum state. 

\end{abstract}

\maketitle


\section{Introduction}

More than a century after the discovery of quantum mechanics, the phenomenon of quantum decoherence, especially in the context of a measurement, remains an unresolved mystery. Some may consider this a solved problem, stating that a measurement is merely an interaction with the surrounding environment, which causes classical dynamics to emerge due to einselection \cite{PhysRevD.24.1516, PhysRevD.26.1862}. 

Almost 30 years later, we now know that there are certain \enquote{measurement} processes that do not disturb the quantum state, these measurements are now known as Quantum Non-Demolition (QND) Measurements \cite{1980Sci...209..547B}. Although these devices still remain limited, one may imagine in the future there will be macroscopic Quantum Non-Demolition Measurement devices, leading us to a \enquote{Schrodinger's Cat} state for a measurement device. As such, the authors believe the process of einselection is not sufficient to fully explain the emergence of classical behaviors in quantum systems. 

There are certain physicists that share this view, and many physicists even believe that the process of quantum decoherence is a key piece of a potential quantum gravity theory. These theories incorporate decoherence through gravitational interactions, and are known as gravitational decoherence \cite{Bassi_2017}. 

In this work, we are not interested in discussing gravitational decoherence, although it is one of the key motivations here. Instead, we consider experiments that may allow us to quantify the conditions in which a state is collapsed. We consider a quantum mechanical system with two entangled components, and change one component in a way that cannot disturb the other in any way except a measurement. This allows us to probe the condition in which the system transitions from quantum to classical, and greatly enhance our understanding of the process of measurement. 

A key tool we use for this experiment is quantum optics. Using processes like spontaneous parametric down-conversion (SPDC) \cite{PhysRevLett.25.84}, we are able to effectively create states with different entangled components. By observing the double slit diffraction pattern in the un-measured component, we are able to determine the conditions in which the system undergoes the quantum to classical transition, making this experiment feasible. 

This work is structured as follows: In section \ref{sec:Motivations}, we outline the importance of the quantifying the classical to quantum transitions. In section \ref{sec:exp_design}, we outline our experiment. In section \ref{sec:diff_pattern}, we calculate the diffraction patterns that will be observed. Lastly, this work concludes with a summary of our findings.

\section{Motivations} \label{sec:Motivations}

Let us consider the following thought experiment. Consider a beam of photons moving through space. These photons meet a device - either a double stilted plate, or a beam splitter. One of the beams meets a measurement device, which absorbs and counts the photon, then re-emits an identical photon. The two beams are then re-directed onto a screen, which observes for double slit interference. 

We may attempt to describe this process quantum mechanically. We can introduce the states $\ket{A}, \ket{B}, \ket{C}$, indicating no photon was detected, photon detected \& device fail respectively. These states are not a single states, but a superposition of many microscopic states that give us the same macroscopic behavior. Since photons are indistinguishable in nature, the re-emitted photons are no different from the absorbed photon. Let us then introduce the states $\ket{0}, \ket{1}, \ket{2}$ for the photons, with $\ket{0}$ indicating that the photon pass through the uninterrupted beam, $\ket{1}$ indicating the photon was absorbed and re-emitted, and $\ket{2}$ indicating it was not re-emitted due to some error. After the beam \enquote{passes through} the measurement device, we then have the superposition:

\begin{align}
    c_1 \ket{A}\ket{0} + c_2 \ket{B}\ket{1} + c_3 \ket{C} \ket{2}.
\end{align}

For a well-designed device, $c_3$ should be sufficiently small. Then, what the above equation tells us is that even if a measurement has been made by the device, there should be still be a double slit patter present on the screen. Yet we did not even require our device be Quantum Non-Demolition in nature - we only required it to have a quantum mechanical description. But, since the measurement is not Quantum Non-Demolition in nature, we should not see an interference patter. One may further imagine putting an observer at the detector, and another observer at the screen. Then, the two observers, based on the knowledge they have of the system, makes incompatible descriptions of the system. These two cannot both be true, thus, we conclude that quantum mechanics cannot be a complete and self-consistent description of nature.

In the consistent histories interpretation of quantum mechanics \cite{consistent_histories}, the same problem is even more visible. In consistent histories, one consider sets of \enquote{histories}. A history is essentially a set of projection onto different basis that happen at different times. In other words, a history of decoherence. In consistent histories, one must know the corresponding history to make predictions of a quantum system. In such a framework, one combines unitary evolution and decoherence. But, for any given quantum system, there may be many alternate \enquote{histories} that are incompatible with each other, like the example we illustrated above. This means that we cannot make any meaningful prediction unless we can predict when the collapse of a wave function will happen - which cannot be done in regular QM. Thus, we are motivated to quantify the conditions that will cause collapse of a wave function. In the later sections, we will detail a potential experiment we think will do exactly that! 


\section{Experimental Design} \label{sec:exp_design}

The authors have though of serval ways to design such an experiment. In the end, we take heavy inspiration from Kim et al.'s \cite{Kim_2000} realization of the delayed choice quantum eraser experiment \cite{PhysRevA.25.2208}. 

\begin{figure}
    \centering
    \includegraphics[width=0.4\textwidth, trim=3cm 5cm 2cm 6cm, clip]{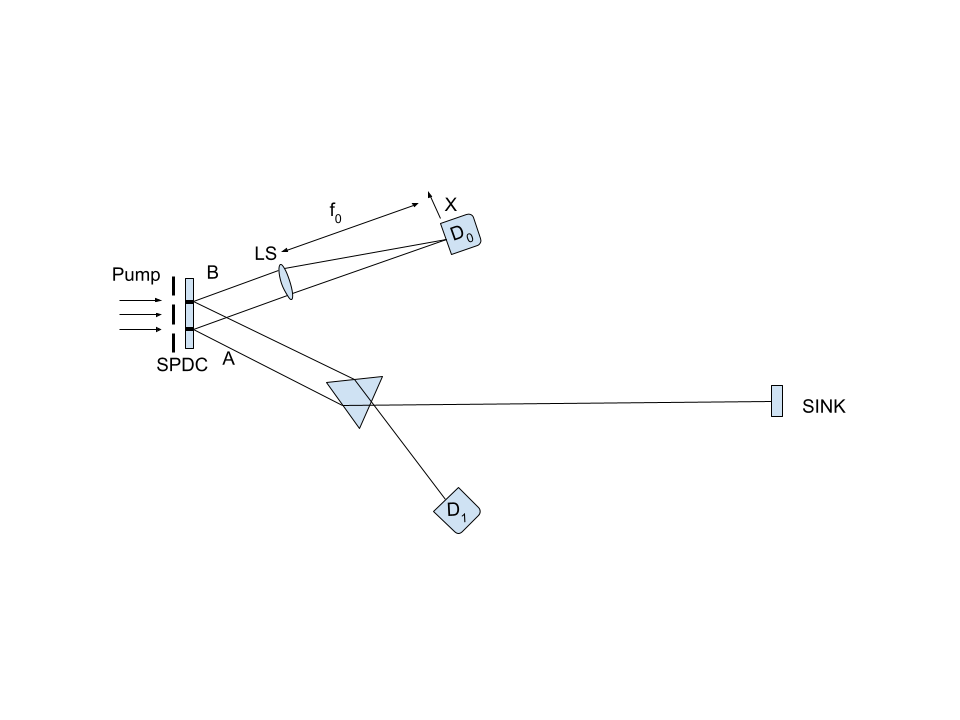} 
    \caption{Proposed Setup of Our Experiment} 
    \label{fig:experiment_diagram}
\end{figure}

We consider a source of photons, and a double slit placed after the source. This source is commonly a pump laser. After the slit, place a degenerate type-II spontaneous parametric down conversion crystal (SPDC). With experimental advances, such crystals with high momentum definition has been reported \cite{PhysRevLett.75.4337, PhysRevA.49.3209}. 
We have drawn the setup of the experiment in figure \ref{fig:experiment_diagram}.

We label the outgoing photons from the slits photon A and photon B, and the outgoing photons from the SPDC idler photon and signal photon. The signal photons from $A$ and $B$ are directed by a lens, so that they will hit detector $D_0$. If the initial photon passes through both slit at once, we will see an interference pattern at detector $D_0$, otherwise, we will not. 

One may worry, will the photon released from the SPDC behave as we expected? Although the SPDC process produces entangled pairs at a rate much lower than the number of incident photons, most of these photons simply establish a suitable boundary condition for the process to occur. For the outgoing photons, the entire process is governed by quantum mechanics, yielding photon pairs (idler and signal) that are entangled and contain \enquote{which-path} information. These photo pairs (idler and signal) are effectively created simultaneously \cite{valencia2004distant}. Furthermore, by examining the classical theory of SPDC in a nonlinear crystal (see, for example, \cite{SPDCTheoryReview}), we find that for each incident photon, the response times are relatively short, and only affected by the speed of light and the optical path length in the crystal. Therefore, when the state is not collapsed, we will expect to observe an interference pattern. 

The idler photon from B is sent outwards and kept interaction free for the duration of the experiment, until it finally hits a sink. This is done by making the optical distance between slit A and the sink much larger than the other optical distances in this experiment. The idler photon from A will be sent to a \enquote{detector} $D_1$. By design, we keep the optical distance between slits A, B and detector $D_0$, and the optical distance between slit B and detector $D_1$ the same. This is so that $D_0$ and $D_1$ will have equal probability of first making a measurement on the \enquote{which-path} information. We are required to do so, to due 2 reasons:
\begin{enumerate}
    \item The wave-packet of the photons having a non-zero size
    \item The \enquote{detectors} we plan to place at $D_1$ will not have timing capabilities 
\end{enumerate}
Since $D_1$ does not have timing capabilities, it will be impossible for the experimenter to determine the sequence of events. To avoid this problem, we make the probability of either happening first equal. Furthermore, due to \enquote{detector} $D_1$ not having timing capabilities, we cannot use joint detection or triggering techniques to separate background noise and signal. Therefore, reduction of environmental noises is needed. This may require the experiment to be operated inside a vacuum and at cryogenic temperatures. 

Notice we use quotations for the \enquote{detector} at $D_1$, as we are interested into placing a variety of objects here to investigate the effects of decoherence, and see what would count as a \enquote{measurement} and cause the superposition at $D_0$ to collapse. This is very much related to the concept of Inherently-Quantum Non-Demolition Measurements that the author proposed in another work \cite{muchenmasterthesis}. In a quantum mechanical description, any local interaction with the idler photons will not affect the signal photons. Thus, any \enquote{measurement} of the idler photon is inherently back-action free in a quantum mechanical description, and any change to the quantum state at $D_0$ must be caused by the collapse of the quantum state. Utilizing this property, we can observe $D_0$ to quantify the parameters that cause the system to collapse. 

Ideally, we start with a \enquote{measurement} process that does not collapse the quantum state, and gradually change certain variables, while keeping others constant, until the states $D_0$ collapses. Following this ideal, we consider 3 different \enquote{detectors} at $D_1$, being:

\begin{enumerate}
    \item Cold Atom Gas
    \item Photo Multiplier Tube
    \item Photographic Plate
\end{enumerate}

And use the different detector to test for the criteria of causing the state at $D_0$ to collapse. In an actual experimental setting, we imagine an experimenter will start with the cold atom gas, and work their way through. This is due to the fact that we believe the cold atom gas is the least likely to collapse the state at $D_0$, and we are very much unsure about the other \enquote{measurement devices}. When performing the actual experiment, the experimenters are very much welcome to change things up following their judgment. 

\subsection{Cold Atom Gases}
We start by considering the cold atom gas. Let us assume we have found a good medium with a absorption spectrum compatible with the idler photons. Then, when a idler photon is absorbed, the cold atom gas has essentially \enquote{learned} the which path information. This is approximately the following process:

\begin{align}
    \ket{1}_{i, B} \ket{0}_{atom} \longrightarrow \ket{0}_{i, B} \ket{1}_{atom},
\end{align}
and one of the key questions here is whether it will collapse the state at detector $D_0$. One may take analogy from the atom-photon interactions in Cavity QED. It has been demonstrated in cavity QED experiments, the atom-light interactions behave quantum mechanically \cite{PhysRevLett.68.1132}, therefore it is reasonable to believe that the photon-atom interaction in our proposed experiment will not collapse the state at $D_0$, presenting a solid platform for the test of subsequent measurement devices. 

\subsection{Photographic Plate}
If the cold atom gas does not cause collapse of the quantum state, we may consider instead placing a photographic plate or film at $D_0$. In the experiment, we leave the plate or film in place, while the readout at $D_0$ is taken continuously. The physical process of happening at the film is as follows:

\begin{align}
    \text{AgX} + \text{light} \rightarrow \text{Ag} + \text{X}^-,
\end{align}

where $AgX$ represents a silver halide crystal. The sliver halide may be unresponsive to a single photon hit, but repeated bombardment, like in an experimental setting, allows the average locations of the photons to be recorded. One may imagine for each hit, one of the silver halide crystal forms a superposition with the which-path information, and does not collapse the state. It is also completely possible that the interaction is significant enough to cause collapse of the quantum state. As a photographic film or plate is often used as a measurement device, for example, to observe a diffraction patter, it is interesting to see what happens with a film placed at $D_1$. 

\subsection{Photo Multiplier Tube}
The photo multiplier tube is the most interesting case. If the previous \enquote{measurements} do not cause the state at $D_0$ to collapse, we may consider using a photomultiplier tube (PMT) to amplify the photons. A photo-multiplier tube utilizes the photo-electric effect to achieve single photon resolution. The physical process goes as follows:

The photon hits a photocathode, which releases an electron (through the photoelectric effect) when the photon hits. A series of dynode is placed afterwards. A voltage is applied, so that the electron accelerates and hits the dynode, and releases more electrons. 

By placing a series of dynodes, the signal is amplified. This amplification is exponential. In a PMT, the total amplification factor is dependent on the number of dynode, or stages of the PMT, and the amplification factor per stage. Let us say that we have an amplification of $G$ per stage, and a total of 5 stages. Then, the total gain by the PMT will be:

\begin{align}
    G_{tot} = G^5.
\end{align}

For the purposes of our proposed experiment, we hope to construct multiple PMTs with special requirements. First, we let our PMT release it's output not onto a readout, but let the electrons fly in free space. We also hope it to have a low amplification factor, and construct multiple PMT's with different number of stages. Then, by switching out the PMT, one is able to quantify exactly the number of stages needed to collapse the  state at $D_0$. By knowing the number of stages, we can determine the number of electrons interacting with the \enquote{which-path} information needed to collapse the state. If performed correctly, this will become the first experiment to observe and quantify the collapse of a quantum state!

\section{Diffraction Patterns at $D_0$} \label{sec:diff_pattern}

To further clarify the experimental setup, let us discuss what will be observed at the detector $D_0$, and demonstrate there is an observable difference when the state is intact verses when it has collapsed. 

Let us consider the optical path from the slit to detector $D_0$ and $D_1$ to be equal. The SPDC process ensures that the idler and signal photon are always created simultaneously \cite{valencia2004distant}. Thus, the temporal distributions of the photons are only affected by the SPDC response time (to the pump photon), and the width of the wave-packet. Experimentally, one may align the response time of the SPDC at slit A and slit B by removing the \enquote{detector} $D_1$ and observing for the double slit interference at $D_0$. When done correctly, the temporal distribution is only affected by the wave packet, and we may ensure that the probability of either the idle photon hitting $D_1$ first and the signal photon hitting $D_0$ first should be equal. 

Let us assume that $D_1$ collapses the quantum state. Then, either $D_1$ detects the photon earlier than $D_0$, or $D_0$ detects the photon earlier than $D_1$. Each of this has probability $50/50$. When the photon is detected first at $D_0$, we expect a interference pattern visible at $D_0$, which is given by:

\begin{align}
    R \propto \mathrm{sinc}^2(\frac{x \pi a}{\lambda f_0}) \cos^2(\frac{x \pi d}{\lambda f_0}),
\end{align}

where $a$ is the length of the slits, $d$ is the distance between the center of the slits, and $f_0$ is the focal length of the lens. If the photon is first detected at $D_1$, we expect no interference patter, so that we have:

\begin{align}
    R \propto \mathrm{sinc}^2(\frac{x \pi a}{\lambda f_0}) .
\end{align}

Therefore, if $D_1$ forms a classical measurement, the interference pattern on the screen takes the form:

\begin{align}
    R \propto \mathrm{sinc}^2(\frac{x \pi a}{\lambda f_0}) (1 + \cos^2(\frac{x \pi d}{\lambda f_0})).
\end{align}

Now, consider when $D_1$ does not form collapse the quantum state. Then, $D_1$ can be thought as a local Hamiltonian operating on the idler photon from B. This is inherently back-action free, i.e., actions on idler photon do not affect the signal photons. Then, it is clear that we expect the following interference pattern at $D_0$:

\begin{align}
    R \propto \mathrm{sinc}^2(\frac{x \pi a}{\lambda f_0}) \cos^2(\frac{x \pi d}{\lambda f_0}),
\end{align}

as the signal photon undergo double-slit interference. The difference in the interference patterns at $D_0$ will allow us to determine exactly under what conditions the states are collapsed.

\section{Conclusion}

In this short article, we propose a method to quantify exactly what conditions will lead to a collapse of a certain quantum state. The experiment utilizes a quantum optics system, and uses photo-multiplier tube with varying stages as it's main variable. By varying the number of stages of the PMT and observing the collapse of the quantum state at the other detector, one may exactly determine how many electrons need to interfere with the "which-path" information to cause collapse of the quantum state. We believe such an experiment is of interest to the physics community, as it is necessary to be able to predict the collapse of a quantum state for a complete and self-consistent quantum theory. We hope this experiment can be performed in the near future.


\bibliography{apssamp}

\providecommand{\noopsort}[1]{}\providecommand{\singleletter}[1]{#1}%
\begin{thebibliography}{14}%
\makeatletter
\providecommand \@ifxundefined [1]{%
 \@ifx{#1\undefined}
}%
\providecommand \@ifnum [1]{%
 \ifnum #1\expandafter \@firstoftwo
 \else \expandafter \@secondoftwo
 \fi
}%
\providecommand \@ifx [1]{%
 \ifx #1\expandafter \@firstoftwo
 \else \expandafter \@secondoftwo
 \fi
}%
\providecommand \natexlab [1]{#1}%
\providecommand \enquote  [1]{``#1''}%
\providecommand \bibnamefont  [1]{#1}%
\providecommand \bibfnamefont [1]{#1}%
\providecommand \citenamefont [1]{#1}%
\providecommand \href@noop [0]{\@secondoftwo}%
\providecommand \href [0]{\begingroup \@sanitize@url \@href}%
\providecommand \@href[1]{\@@startlink{#1}\@@href}%
\providecommand \@@href[1]{\endgroup#1\@@endlink}%
\providecommand \@sanitize@url [0]{\catcode `\\12\catcode `\$12\catcode `\&12\catcode `\#12\catcode `\^12\catcode `\_12\catcode `\%12\relax}%
\providecommand \@@startlink[1]{}%
\providecommand \@@endlink[0]{}%
\providecommand \url  [0]{\begingroup\@sanitize@url \@url }%
\providecommand \@url [1]{\endgroup\@href {#1}{\urlprefix }}%
\providecommand \urlprefix  [0]{URL }%
\providecommand \Eprint [0]{\href }%
\providecommand \doibase [0]{https://doi.org/}%
\providecommand \selectlanguage [0]{\@gobble}%
\providecommand \bibinfo  [0]{\@secondoftwo}%
\providecommand \bibfield  [0]{\@secondoftwo}%
\providecommand \translation [1]{[#1]}%
\providecommand \BibitemOpen [0]{}%
\providecommand \bibitemStop [0]{}%
\providecommand \bibitemNoStop [0]{.\EOS\space}%
\providecommand \EOS [0]{\spacefactor3000\relax}%
\providecommand \BibitemShut  [1]{\csname bibitem#1\endcsname}%
\let\auto@bib@innerbib\@empty
\bibitem [{\citenamefont {Zurek}(1981)}]{PhysRevD.24.1516}%
  \BibitemOpen
  \bibfield  {author} {\bibinfo {author} {\bibfnamefont {W.~H.}\ \bibnamefont {Zurek}},\ }\bibfield  {title} {\bibinfo {title} {Pointer basis of quantum apparatus: Into what mixture does the wave packet collapse?},\ }\href {https://doi.org/10.1103/PhysRevD.24.1516} {\bibfield  {journal} {\bibinfo  {journal} {Phys. Rev. D}\ }\textbf {\bibinfo {volume} {24}},\ \bibinfo {pages} {1516} (\bibinfo {year} {1981})}\BibitemShut {NoStop}%
\bibitem [{\citenamefont {Zurek}(1982)}]{PhysRevD.26.1862}%
  \BibitemOpen
  \bibfield  {author} {\bibinfo {author} {\bibfnamefont {W.~H.}\ \bibnamefont {Zurek}},\ }\bibfield  {title} {\bibinfo {title} {Environment-induced superselection rules},\ }\href {https://doi.org/10.1103/PhysRevD.26.1862} {\bibfield  {journal} {\bibinfo  {journal} {Phys. Rev. D}\ }\textbf {\bibinfo {volume} {26}},\ \bibinfo {pages} {1862} (\bibinfo {year} {1982})}\BibitemShut {NoStop}%
\bibitem [{\citenamefont {{Braginskii}}\ \emph {et~al.}(1980)\citenamefont {{Braginskii}}, \citenamefont {{Vorontsov}},\ and\ \citenamefont {{Thorne}}}]{1980Sci...209..547B}%
  \BibitemOpen
  \bibfield  {author} {\bibinfo {author} {\bibfnamefont {V.~B.}\ \bibnamefont {{Braginskii}}}, \bibinfo {author} {\bibfnamefont {I.~I.}\ \bibnamefont {{Vorontsov}}},\ and\ \bibinfo {author} {\bibfnamefont {K.~S.}\ \bibnamefont {{Thorne}}},\ }\bibfield  {title} {\bibinfo {title} {{Quantum Nondemolition Measurements}},\ }\href {https://doi.org/10.1126/science.209.4456.547} {\bibfield  {journal} {\bibinfo  {journal} {Science}\ }\textbf {\bibinfo {volume} {209}},\ \bibinfo {pages} {547} (\bibinfo {year} {1980})}\BibitemShut {NoStop}%
\bibitem [{\citenamefont {Bassi}\ \emph {et~al.}(2017)\citenamefont {Bassi}, \citenamefont {Großardt},\ and\ \citenamefont {Ulbricht}}]{Bassi_2017}%
  \BibitemOpen
  \bibfield  {author} {\bibinfo {author} {\bibfnamefont {A.}~\bibnamefont {Bassi}}, \bibinfo {author} {\bibfnamefont {A.}~\bibnamefont {Großardt}},\ and\ \bibinfo {author} {\bibfnamefont {H.}~\bibnamefont {Ulbricht}},\ }\bibfield  {title} {\bibinfo {title} {Gravitational decoherence},\ }\href {https://doi.org/10.1088/1361-6382/aa864f} {\bibfield  {journal} {\bibinfo  {journal} {Classical and Quantum Gravity}\ }\textbf {\bibinfo {volume} {34}},\ \bibinfo {pages} {193002} (\bibinfo {year} {2017})}\BibitemShut {NoStop}%
\bibitem [{\citenamefont {Burnham}\ and\ \citenamefont {Weinberg}(1970)}]{PhysRevLett.25.84}%
  \BibitemOpen
  \bibfield  {author} {\bibinfo {author} {\bibfnamefont {D.~C.}\ \bibnamefont {Burnham}}\ and\ \bibinfo {author} {\bibfnamefont {D.~L.}\ \bibnamefont {Weinberg}},\ }\bibfield  {title} {\bibinfo {title} {Observation of simultaneity in parametric production of optical photon pairs},\ }\href {https://doi.org/10.1103/PhysRevLett.25.84} {\bibfield  {journal} {\bibinfo  {journal} {Phys. Rev. Lett.}\ }\textbf {\bibinfo {volume} {25}},\ \bibinfo {pages} {84} (\bibinfo {year} {1970})}\BibitemShut {NoStop}%
\bibitem [{\citenamefont {Griffiths}(1984)}]{consistent_histories}%
  \BibitemOpen
  \bibfield  {author} {\bibinfo {author} {\bibfnamefont {R.}~\bibnamefont {Griffiths}},\ }\bibfield  {title} {\bibinfo {title} {Consistent histories and the interpretation of quantum mechanics},\ }\href {https://doi.org/10.1007/BF01015734} {\bibfield  {journal} {\bibinfo  {journal} {Journal of Statistical Physics}\ }\textbf {\bibinfo {volume} {36}},\ \bibinfo {pages} {219} (\bibinfo {year} {1984})}\BibitemShut {NoStop}%
\bibitem [{\citenamefont {Kim}\ \emph {et~al.}(2000)\citenamefont {Kim}, \citenamefont {Yu}, \citenamefont {Kulik}, \citenamefont {Shih},\ and\ \citenamefont {Scully}}]{Kim_2000}%
  \BibitemOpen
  \bibfield  {author} {\bibinfo {author} {\bibfnamefont {Y.-H.}\ \bibnamefont {Kim}}, \bibinfo {author} {\bibfnamefont {R.}~\bibnamefont {Yu}}, \bibinfo {author} {\bibfnamefont {S.~P.}\ \bibnamefont {Kulik}}, \bibinfo {author} {\bibfnamefont {Y.}~\bibnamefont {Shih}},\ and\ \bibinfo {author} {\bibfnamefont {M.~O.}\ \bibnamefont {Scully}},\ }\bibfield  {title} {\bibinfo {title} {Delayed “choice” quantum eraser},\ }\href {https://doi.org/10.1103/physrevlett.84.1} {\bibfield  {journal} {\bibinfo  {journal} {Physical Review Letters}\ }\textbf {\bibinfo {volume} {84}},\ \bibinfo {pages} {1–5} (\bibinfo {year} {2000})}\BibitemShut {NoStop}%
\bibitem [{\citenamefont {Scully}\ and\ \citenamefont {Dr\"uhl}(1982)}]{PhysRevA.25.2208}%
  \BibitemOpen
  \bibfield  {author} {\bibinfo {author} {\bibfnamefont {M.~O.}\ \bibnamefont {Scully}}\ and\ \bibinfo {author} {\bibfnamefont {K.}~\bibnamefont {Dr\"uhl}},\ }\bibfield  {title} {\bibinfo {title} {Quantum eraser: A proposed photon correlation experiment concerning observation and "delayed choice" in quantum mechanics},\ }\href {https://doi.org/10.1103/PhysRevA.25.2208} {\bibfield  {journal} {\bibinfo  {journal} {Phys. Rev. A}\ }\textbf {\bibinfo {volume} {25}},\ \bibinfo {pages} {2208} (\bibinfo {year} {1982})}\BibitemShut {NoStop}%
\bibitem [{\citenamefont {Kwiat}\ \emph {et~al.}(1995)\citenamefont {Kwiat}, \citenamefont {Mattle}, \citenamefont {Weinfurter}, \citenamefont {Zeilinger}, \citenamefont {Sergienko},\ and\ \citenamefont {Shih}}]{PhysRevLett.75.4337}%
  \BibitemOpen
  \bibfield  {author} {\bibinfo {author} {\bibfnamefont {P.~G.}\ \bibnamefont {Kwiat}}, \bibinfo {author} {\bibfnamefont {K.}~\bibnamefont {Mattle}}, \bibinfo {author} {\bibfnamefont {H.}~\bibnamefont {Weinfurter}}, \bibinfo {author} {\bibfnamefont {A.}~\bibnamefont {Zeilinger}}, \bibinfo {author} {\bibfnamefont {A.~V.}\ \bibnamefont {Sergienko}},\ and\ \bibinfo {author} {\bibfnamefont {Y.}~\bibnamefont {Shih}},\ }\bibfield  {title} {\bibinfo {title} {New high-intensity source of polarization-entangled photon pairs},\ }\href {https://doi.org/10.1103/PhysRevLett.75.4337} {\bibfield  {journal} {\bibinfo  {journal} {Phys. Rev. Lett.}\ }\textbf {\bibinfo {volume} {75}},\ \bibinfo {pages} {4337} (\bibinfo {year} {1995})}\BibitemShut {NoStop}%
\bibitem [{\citenamefont {Kwiat}\ \emph {et~al.}(1994)\citenamefont {Kwiat}, \citenamefont {Eberhard}, \citenamefont {Steinberg},\ and\ \citenamefont {Chiao}}]{PhysRevA.49.3209}%
  \BibitemOpen
  \bibfield  {author} {\bibinfo {author} {\bibfnamefont {P.~G.}\ \bibnamefont {Kwiat}}, \bibinfo {author} {\bibfnamefont {P.~H.}\ \bibnamefont {Eberhard}}, \bibinfo {author} {\bibfnamefont {A.~M.}\ \bibnamefont {Steinberg}},\ and\ \bibinfo {author} {\bibfnamefont {R.~Y.}\ \bibnamefont {Chiao}},\ }\bibfield  {title} {\bibinfo {title} {Proposal for a loophole-free bell inequality experiment},\ }\href {https://doi.org/10.1103/PhysRevA.49.3209} {\bibfield  {journal} {\bibinfo  {journal} {Phys. Rev. A}\ }\textbf {\bibinfo {volume} {49}},\ \bibinfo {pages} {3209} (\bibinfo {year} {1994})}\BibitemShut {NoStop}%
\bibitem [{\citenamefont {Valencia}\ \emph {et~al.}(2004)\citenamefont {Valencia}, \citenamefont {Scarcelli},\ and\ \citenamefont {Shih}}]{valencia2004distant}%
  \BibitemOpen
  \bibfield  {author} {\bibinfo {author} {\bibfnamefont {A.}~\bibnamefont {Valencia}}, \bibinfo {author} {\bibfnamefont {G.}~\bibnamefont {Scarcelli}},\ and\ \bibinfo {author} {\bibfnamefont {Y.}~\bibnamefont {Shih}},\ }\bibfield  {title} {\bibinfo {title} {Distant clock synchronization using entangled photon pairs},\ }\href@noop {} {\bibfield  {journal} {\bibinfo  {journal} {Applied Physics Letters}\ }\textbf {\bibinfo {volume} {85}},\ \bibinfo {pages} {2655} (\bibinfo {year} {2004})}\BibitemShut {NoStop}%
\bibitem [{\citenamefont {Couteau}(2018)}]{SPDCTheoryReview}%
  \BibitemOpen
  \bibfield  {author} {\bibinfo {author} {\bibfnamefont {C.}~\bibnamefont {Couteau}},\ }\bibfield  {title} {\bibinfo {title} {Spontaneous parametric down-conversion},\ }\href {https://doi.org/10.1080/00107514.2018.1488463} {\bibfield  {journal} {\bibinfo  {journal} {Contemporary Physics}\ }\textbf {\bibinfo {volume} {59}},\ \bibinfo {pages} {291} (\bibinfo {year} {2018})},\ \Eprint {https://arxiv.org/abs/https://doi.org/10.1080/00107514.2018.1488463} {https://doi.org/10.1080/00107514.2018.1488463} \BibitemShut {NoStop}%
\bibitem [{\citenamefont {He}(2024)}]{muchenmasterthesis}%
  \BibitemOpen
  \bibfield  {author} {\bibinfo {author} {\bibfnamefont {M.}~\bibnamefont {He}},\ }\emph {\bibinfo {title} {Investigating Quantum Non-Demolition Measurements with Cavity Optomechanics: Theory and Experimental Approaches}},\ \href@noop {} {Master's thesis},\ \bibinfo  {school} {Imperial College London} (\bibinfo {year} {2024}),\ \bibinfo {note} {forthcoming}\BibitemShut {NoStop}%
\bibitem [{\citenamefont {Thompson}\ \emph {et~al.}(1992)\citenamefont {Thompson}, \citenamefont {Rempe},\ and\ \citenamefont {Kimble}}]{PhysRevLett.68.1132}%
  \BibitemOpen
  \bibfield  {author} {\bibinfo {author} {\bibfnamefont {R.~J.}\ \bibnamefont {Thompson}}, \bibinfo {author} {\bibfnamefont {G.}~\bibnamefont {Rempe}},\ and\ \bibinfo {author} {\bibfnamefont {H.~J.}\ \bibnamefont {Kimble}},\ }\bibfield  {title} {\bibinfo {title} {Observation of normal-mode splitting for an atom in an optical cavity},\ }\href {https://doi.org/10.1103/PhysRevLett.68.1132} {\bibfield  {journal} {\bibinfo  {journal} {Phys. Rev. Lett.}\ }\textbf {\bibinfo {volume} {68}},\ \bibinfo {pages} {1132} (\bibinfo {year} {1992})}\BibitemShut {NoStop}%
\end{thebibliography}%


\providecommand{\noopsort}[1]{}\providecommand{\singleletter}[1]{#1}%
%

\end{document}